\documentclass{article}

\usepackage[final,nonatbib]{neurips_2021}
\usepackage{enumitem}

\usepackage[utf8]{inputenc} % allow utf-8 input
\usepackage[T1]{fontenc}    % use 8-bit T1 fonts
\usepackage{hyperref}       % hyperlinks
\usepackage{url}            % simple URL typesetting
\usepackage{booktabs}       % professional-quality tables
\usepackage{amsfonts}       % blackboard math symbols
\usepackage{nicefrac}       % compact symbols for 1/2, etc.
\usepackage{microtype}      % microtypography
\usepackage{xcolor}         % colors

\usepackage{graphicx}       % figure
\usepackage{biblatex}       % reference
\usepackage{algorithm2e}    % pseudocode
\title{Engineering AI Tools for Systematic and Scalable Quality Assessment in Magnetic Resonance Imaging}

\author{%
  Yukai~Zou\thanks{Equal Contribution} \\
  University of Southampton\\
  University Hospital Southampton NHS Foundation Trust\\
  \texttt{Y.Zou@soton.ac.uk} \\
   \And
   Ikbeom~Jang\footnotemark[1] \\
   Massachusetts General Hospital \\
   Harvard Medical School \\
   \texttt{ikbeom.jang@mgh.harvard.edu} \\
}

\begin{document}

\maketitle

\begin{abstract}
  A desire to achieve large medical imaging datasets keeps increasing as machine learning algorithms, parallel computing, and hardware technology evolve. Accordingly, there is a growing demand in pooling data from multiple clinical and academic institutes to enable large-scale clinical or translational research studies. Magnetic resonance imaging (MRI) is a frequently used, non-invasive imaging modality. However, constructing a big MRI data repository has multiple challenges related to privacy, data size, DICOM format, logistics, and non-standardized images. Not only building the data repository is difficult, but using data pooled from the repository is also challenging, due to heterogeneity in image acquisition, reconstruction, and processing pipelines across MRI vendors and imaging sites. This position paper describes challenges in constructing a large MRI data repository and using data downloaded from such data repositories in various aspects. To help address the challenges, the paper proposes introducing a quality assessment pipeline, with considerations and general design principles.
\end{abstract}

\section{Background}

A desire to achieve a large medical imaging dataset keeps increasing as machine learning algorithms, parallel computing, and hardware evolve. Accordingly, there is a growing desire for larger datasets by pooling harmonized data from multiple institutions. One of the most frequently used imaging modalities is Magnetic Resonance Imaging (MRI). MRI can non-invasively monitor the physiological processes of the human body under health and disease conditions. There is no ionizing radiation compared to X-ray Computed Tomography (CT), allowing subjects to be monitored at multiple time points without the concern of tissue damage due to radiation exposure.

In T1- and T2-weighted imaging, MR signals are stored in relatively arbitrary values, and this provides anatomical contrasts that radiologist is traditionally trained to read and make clinical decisions. With the advancement of hardware and infrastructure for data collection, storage, and computing, the development of quantitative MRI techniques becomes fast-evolving and adds new dimensions to probing normal biological processes and pathophysiology. In quantitative MRI, parameters are extracted from the MRI signals through computational modeling, which can serve as an imaging-based biomarker that is objectively measured and evaluated to correspond to a specific physical or physiological property \cite{Biomarker2001}. Therefore, quantitative MRI opens up a new way of thinking about MRI signals, imaging processes, optimization tools, and scientific computing, spanning from physiology to tissue microstructure and morphology, and the research activities are highly interdisciplinary \cite{Seiberlich2020}.

However, the interpretation of imaging-based biomarkers still faces barriers. Clinical translation of quantitative MRI has been slow, with cautions about directly using the pooled data. The barriers of \textit{poor reproducibility} and \textit{low efficiency} need to be overcome in order to translate quantitative MRI into clinical practice. As such, AI tools hold the potential to fulfill the unmet need so that high-quality MRI data can be guaranteed to be integrated into routine clinical care.

\section{Challenges} 

% Section 2.1
\subsection{Challenges in constructing multi-site MRI data repositories} 

Constructing a large MRI data repository has multiple challenges. MRI data are stored as ``Digital Imaging and Communications in Medicine (DICOM)", a standardized format for the storage, exchange, and transmission of medical imaging data. A DICOM consists of a header and the actual image itself. The header consists of metadata that describes the image, acquisition parameters, study and participant information, institution, etc. Once raw MRI data are acquired, the data are reconstructed into images that can be clinically interpreted. To note, the raw data are typically not stored as DICOM, given that they are often huge in file size.

% File size
DICOM files are typically large in size, which can be an issue because both the data storage and transfer come at a price. For example, it takes about 1--1.5 GB of storage to host the data for one routine exam of a clinical brain study. Often, each patient has multiple exams, i.e., data at multiple time points. About 6--13.5 TB are required when building a data repository for a longitudinal study that comprises 3 timepoints, about 1,000 patients, with 1--2 backups. This is even without adding 4-D MRI data, such as functional or diffusion MRI. One potential solution can be to utilize image compression technology. Recent technologies using generative adversarial networks make it possible to compress images by a factor of up to 100 without significantly compromising image quality \cite{mentzer2020high}.

% Privacy & de-identification
DICOMs must be de-identified before uploads. Because DICOM files contain metadata that include protected health information (PHI) and personally identifiable information (PII), it is considered a violation to Health Insurance Portability and Accountability Act (HIPAA) to access, disclose, or share them with purposes other than treatment, healthcare operation, or payment unless an authorization is obtained from the patient in advance. In principle, PHI in a DICOM header may be de-identified in two ways \cite{aryanto2015free}. The \textit{anonymization} removes information carried by header elements or replaces it with random values so that the original identity cannot be restored at all. The \textit{pseudonymization} is a more frequently used method that replaces the real identifiers within a data record using an artificial identifier and a list of mappings, that could be used by authorized personnel to track down the original identity of the patient \cite{neubauer2011methodology, neubauer2008improving, noumeir2007pseudonymization}. Not only DICOM header entries but the image itself is advised to be anonymized because facial features can be rendered from the image. Defacing algorithms (e.g., \cite{bischoff2007technique}) have been developed that use models of non-brain structures for removing facial features that may potentially allow the identification of a participant from their MR scan.

% Logistics: Alternative to constructing data repository (->federated learning)
To build a big MRI data repository, hospitals and other institutes need to collaborate and host centralized databases. This overhead can quickly become a logistical challenge and usually requires time-consuming approval processes due to data privacy and ethical concerns associated with data sharing in healthcare \cite{larson2020ethics}. Even when these challenges can be addressed, institutions owning the valuable dataset may prefer not to make it publicly available \cite{roth2020federated}. One alternative to the data sharing hurdles is federated learning \cite{mcmahan2017communication}, where only model weights are shared between participating institutions without sharing the actual data.

% Data retrieval from clinical systems
Data retrieval from clinical systems also poses some challenges. Retrieving large amounts of data may cause issues that lead to a breakdown of the Picture Archiving and Communication System (PACS), which can be detrimental for clinicians and technicians. Cloud services such as Amazon Web Services and Microsoft Azure can provide solutions that ensure secure and scalable storage, as well as fast and secure data transmission. However, when switching to a cloud-based solution, the risk of data retrieval must be well managed. For security purposes, the types of data that cannot be hosted on a public repository must be identified beforehand; in addition, a strategy such as \textit{prefetching} \cite{hayotsasson2021benefits} allows marking those data pulled from remote repositories to be deleted after the analysis is done.

% Section 2.2
\subsection{Challenges in using data pooled from MRI data repositories} 

% Heterogeneity in MRI data
Using data downloaded from such data repositories can also be challenging in several aspects. First, directly using the pooled data is not advised, due to heterogeneity of the MRI data in acquisition parameters, magnetic field strength, receiver coil, reconstruction parameters, post-processing parameters, and operators who acquired the data. The images acquired are not standardized across the major vendors of clinical MRI machines, including GE, Philips, Siemens, and Canon, which makes data harmonization particularly challenging.

% Image quality
Image quality is another factor to consider. Adding data with poor quality often harms the model, hence it is better to filter them out. For this reason, many hospitals and institutes have a quality assessment pipeline that filters out corrupted data or data with too much noise or artifacts.

% DICOM
Although DICOM has already been standardized to a large extent to enable medical imaging data more accessible, there are numerous private DICOM tags that are defined by different vendors. Those private tags often contain important parameters, e.g., coil element, acquisition, and storage of the data, etc. Unfortunately, the naming of those tags is often not descriptive, causing difficulty in parsing and filtering the data. Several open-source software provides functionality to handle these private tags -- e.g., \textit{dicom\_parser} for Siemens tags \cite{dicomparser2021}.

\begin{figure}[bhtp]
  \centering
  %\fbox{\rule[-.5cm]{0cm}{4cm} \rule[-.5cm]{4cm}{0cm}}
  \includegraphics[width=1.0\linewidth]{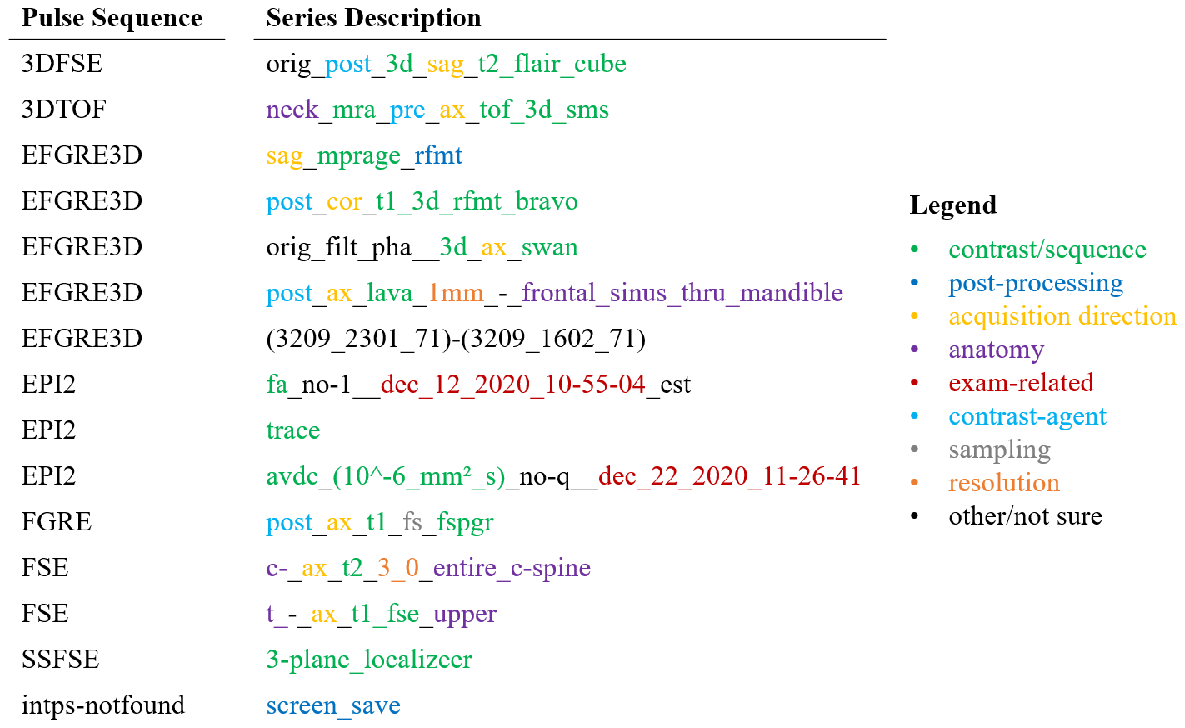}
  \caption{Highly variable MRI ‘Series Description’s. Above are a few examples from a clinical GE scanner.}
\label{fig:series_desc}
\end{figure}

% Series selection (moved from 2.1)
Series selection from MRI exams can be extremely challenging. The most informative DICOM tag related to series selection is ``Series Description", however, there is no standard form to describe a series and mixes information of different nature such as workflow, acquisition, anatomy, and post-processing, e.g., Ax T1 Pre whole brain, Sag MPRAGE RFMT. Also, a wide variety of non-standard annotation formats are used and they are not inter-operable. See Fig. \ref{fig:series_desc} for a few more examples.

% Memory and file size
Another major difficulty in using a large MRI dataset is that it takes too much memory, especially when housing large quantities of 3D data. Training a model (e.g., deep learning model) with such large data can be troublesome because one may not be able to fit the data into the limited graphic card memory and would have to write a project-specific data loader.

\section{Quality Assessment (QA): Considerations and General Design Principles} 

Introducing a QA pipeline provides a way to resolve at least some of the challenges described above by ensuring important elements for a certain study. Such elements not only include image quality but also can include acquisition parameters, processing pipeline, sequence, and anatomy.

The ways ``quality" is defined and assessed have been well established within the world of business. Quality refers to the perception of the degree to which a set of characteristics of the product or service meets the customer's expectations. QA, according to Mainz et al. \cite{Mainz1992}, is ``the data collection and analysis through which the degree of conformity to predetermined standards and criteria are exemplified." When the product or service has been found unsatisfactory through QA, actions shall be made to identify the cause and to make improvements \cite{Mainz1992}.

\begin{figure}[bhtp]
  \centering
  %\fbox{\rule[-.5cm]{0cm}{4cm} \rule[-.5cm]{4cm}{0cm}}
  \includegraphics[width=1.0\linewidth]{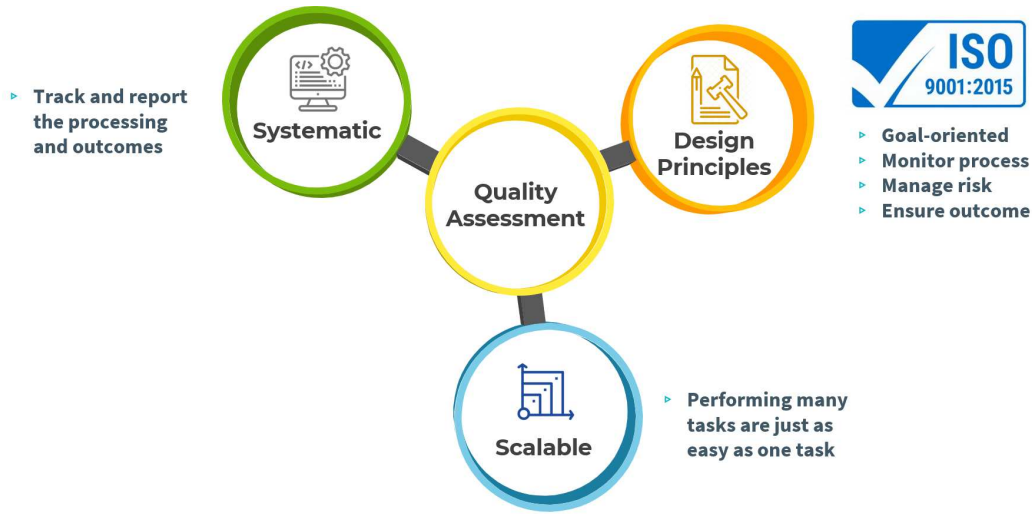}
  \caption{Considerations and general design principles to build a quality assessment pipeline.}
\label{fig:considerations}
\end{figure}

QA for MRI needs to be both \textit{systematic} and \textit{scalable} (Fig. \ref{fig:considerations}). The complexities of MRI processing require a procedure that is capable of systematically tracking and reporting the processing and outcomes. It needs to take into account what steps are involved in the image processing, the input and output of each step, the purpose of the QA, and whether the procedure spans multiple institutes. The QA procedure should also be scalable, meaning that performing hundreds and thousands of QA should be as easy and efficient as performing only one QA. Addressing these factors will significantly make it faster for the research community to use the datasets without concerning the quality.

General design principles of establishing a quality management system apply for engineering AI tools for systematic and scalable QA for MRI. According to the International Organization for Standardization (ISO9001:2015) \cite{ISO9001:2015}, there are four principles (Fig. \ref{fig:considerations}):

\begin{enumerate}[leftmargin=0.4cm]
    \item \textbf{Goal-oriented.} It is important to keep in mind who the customer is, and depending on the needs of the customer, the set of characteristics for evaluating image quality can be different, hence the goals would vary. For radiologists who read images and identify regions with pathology, QA should focus on the image quality such as contrast and artifacts. For researchers who carry out big data analysis, the goal would be to provide a dataset that have been properly filtered and harmonized.
    
    \item \textbf{Monitor process.} To develop reliable AI tools in the medical field, the following items are recommended to be monitored: reliability of the reference standards, whether the training dataset matches the intended use, independence of hyperparameter tuning, and independence of model evaluation \cite{Hagiwara2020}. Because the QA pipeline for MRI can comprise numerous steps, key elements within each step need to be listed, with standardized operation procedures.
    
    \item \textbf{Manage risk.} According to Ropeik and Gray \cite{ropeik2002risk}, there are four elements to consider: \textit{hazard}, \textit{exposure}, \textit{consequence}, and \textit{probability}. What is the probability that exposure to hazard will cause a negative consequence? This question can be asked throughout the QA pipeline, from acquisition to reconstruction and to processing, to ensure MRI data are reliable for accurate patient diagnoses and real-world applications \cite{edupuganti2020uncertainty}.
    
    \item \textbf{Ensure outcomes.} Addressing the issue of reproducibility \cite{Poldrack2017a} is the key to ensuring reliable outcomes. As researchers that strive for best practices in neuroscience, we ignore reproducibility at our own peril \cite{POLDRACK201911}. AI tools can effectively eliminate the primary source of variability, which is fatigue from manual labeling. For quantitative MRI, ensuring reproducible objective measures require such AI tools can unfold the complexity in data acquisition and post-processing.
\end{enumerate}

% \section*{References}
\medskip
{
\small
\printbibliography
}
%\bibliography{references}
%\bibliographystyle{IEEEtran}

%%%%%%%%%%%%%%%%%%%%%%%%%%%%%%%%%%%%%%%%%%%%%%%%%%%%%%%%%%%%

% \appendix

% \section{Appendix}

% Optionally include extra information (complete proofs, additional experiments and plots) in the appendix.
% This section will often be part of the supplemental material.

\end{document}